\documentclass[11pt]{article}
\usepackage[margin=2.5cm]{geometry}
\usepackage[utf8]{inputenc}
\usepackage{amsmath,amsthm}    
\usepackage{hyperref}
\usepackage{amsfonts}
\usepackage{graphicx}
\usepackage{colortbl}
\usepackage{multirow}
\usepackage{array}
\usepackage{authblk}
\usepackage{todonotes}
\usepackage{lineno}
\usepackage{float}
\usepackage{graphicx}
\usepackage{color}
\usepackage{subfig}
\usepackage{tcolorbox}
\usepackage{bm}
\usepackage{mathtools}
\usepackage{commath}
\usepackage[detect-none]{siunitx}
\usepackage[sort&compress,numbers]{natbib}

\usepackage{array}
\newcolumntype{L}[1]{>{\raggedright\let\newline\\\arraybackslash\hspace{0pt}}m{#1}}
\newcolumntype{C}[1]{>{\centering\let\newline\\\arraybackslash\hspace{0pt}}m{#1}}
\newcolumntype{R}[1]{>{\raggedleft\let\newline\\\arraybackslash\hspace{0pt}}m{#1}}

\newtheorem{theorem}{Theorem}[section]
\newtheorem{lemma}{Lemma}[section]

\newtheorem*{remark}{Remark}

\def\r0{\mathcal{R}_0}

\title{\textbf{Prediction of vaccination coverage level in the heterogeneous mixing population}}
\author[1,2]{Fan Bai\thanks{fanbai@uic.edu.cn}}
\author[1]{Qianyu Chen\thanks{q030033002@mail.uic.edu.cn}}
\author[1]{Yizhuo Xu\thanks{q030033048@mail.uic.edu.cn}}
\affil[1]{Department of Mathematical Sciences, \protect \\
	Faculty of Science and Technology, \protect\\ Beijing Normal University - Hong Kong Baptist University United International College, \protect\\ Zhuhai, Guangdong, China}

\affil[2]{Guangdong Provincial Key Laboratory of Interdisciplinary Research and
	Application for Data Science, \protect \\
	BNU-HKBU United International College, \protect \\ 
	Zhuhai, Guangdong, China}

\begin{document}
\maketitle

\medskip

\\

\medskip


\section{Abstract}
Heterogeneity of population is a key factor in modeling the transmission of disease among the population and has huge impact on the outcome of the transmission. In order to investigate the decision making process in the heterogeneous mixing population regarding whether to be vaccinated or not, we propose the modeling framework which includes the epidemic models and the game theoretical analysis. We consider two sources of heterogeneity in this paper: the different activity levels and the different relative vaccination costs. It is interesting to observe that, if both sources of heterogeneity are considered, there exist a finite number of Nash equilibria (evolutionary stable strategies (ESS)) of the vaccination game. While if only the difference of activity levels is considered, there are infinitely many Nash equilibira. For the latter case, the initial condition of the decision making process becomes highly sensitive. In the application of public health management, the inclusion of population heterogeneity significantly complicates the prediction of the overall vaccine coverage level.

\medspace

\section{Introduction}
\subsection{Vaccination games and predicted vaccine coverage levels}

Vaccination is one of the most effective methods for prevention of the spread of infectious diseases. Based on the principle that every individual in the population is trying to maximize the own payoff, all individuals in the population are supposed to make rational decisions on whether or not to take vaccination. Ultimately, a certain percentage of individuals determines to take the vaccine, while the rest determine to be non-vaccinated. This leads to the overall vaccine coverage level in the population. In this paper, we mainly focus on the process of decision making regarding vaccination and consider the process is one type of evolution but with a shorter time scale. Thus, evolutionary game theory is appropriate to be applied to study the dynamics of this process. The first paper to discuss the connection between game theory and epidemiology was written by Fine and Clarkson in $1986$ (see \citep{Fine1986}). Subsequently, several papers discussed vaccination games in infinite population, \textit{e}.\textit{g}., \citep{Bauch2004, Tassier2013, Bai2016, Bai2019,Galvani2007,Reluga2006,Shim2012,Shim2012a}. It has been investigated in papers \cite{Bai2016,Bai2016a}, the Nash equilibrium of the vaccination game is unique if the population under study is homogeneous mixing. The investigation was extended to study the population with heterogeneous mixing in paper \cite{Bai2020}. In \cite{Bai2020}, the different activity levels for sub-populations in the wholly population is the only source of the heterogeneity. It was proved that, the Nash equilibrium of the vaccination game in the heterogeneous mixing population is not unique. Another significant contribution was to show that, even the heterogeneity is introduced into the game theoretical analysis, the herd immunity for the population can not be achieved if all individuals are driven by self interest. The final fraction of vaccinated individuals is therefore sub-optimal. However, one of the major factors which affects the decision-making process is the difference of relative vaccination costs for all sub-populations and it was not investigated. In this paper, we make the general assumption that the occurrence of decision making process is prior to the outbreak of the disease. Information about possible outcomes of infection is provided to all individuals, and all individuals use same information to assess possible outcomes (\citep{Bai2016a,Bauch2004}). We consider the heterogeneity which arise from two main sources: the different activity levels and the varied relative vaccination costs. The general formulation of modeling the decision making process in the heterogeneous mixing population is introduced, analyzed and simulated. The impact of transmission of infection is ignored, the approach of differential games was discuss in \citep{Reluga2010} and other papers.

\medspace

\subsection{The impact of heterogeneity}
To realistically simulate the spread of diseases in the population, it is critical to take into account the structure of the population (\cite{Inaba2017}). Some factors to cause the heterogeneity in the population include the chronological age, the activity level and the immunity level. The structured epidemic models are already well studied, we refer \cite{Auger2008,Inaba2017,Brauer2008book} for detailed discussions regarding this topic. The major novelty of this paper is to consider the varied relative costs of the vaccination in different sub-populations as one type of source of the heterogeneity. Throughout this paper, we define the relative cost of vaccination as the ratio of the cost of vaccination and the cost of the infection for a certain type of disease. With the combination of mathematical modeling for structured populations and game theoretical approach, it is shown in this paper that the heterogeneity originated from different relative costs of vaccination dramatically change the dynamics of the whole decision making process. There exist multiple Nash equilibria or evolutionary stable strategies (ESS) for the vaccination games and some of the Nash equilibira are locally stable. This conclusion can be applied to predict the vaccine coverage level in the structured population.

\medskip

\subsection{Outline}
The paper consists of three parts. In the first part, we formulate the general mathematical modeling framework in order to predict the vaccine coverage level in the structured population. We first propose the age-of-infection mathematical model which consists of integro-differential equations for the considered structured population, then calculate the basic reproduction number $\mathcal{R}_0$ and derive the final size relation, which directly leads to the computations of attack ratios for all sub-populations. Attack ratios are expressed as functions of vaccine coverage levels and will be used in the game theoretical analysis. The second ingredient of the modeling framework is to apply the vaccination game in the heterogeneous mixing population. Without the loss of generality, we consider the vaccination game in the population with two sub-groups, the extension to the case with more than two sub-groups is direct. In the second part of the paper, we apply the method of replicator equations to analyze the models we have formulated. The dynamics of the decision making process is driven by the changes of individuals' payoffs in different sub-populations. The equilibria or the stable states of the model are computed, which represent the possible final outcome of the decision making process. In the third part, we perform a series of numerical experiments to simulate the decision making process we have described. The parameters which represent the heterogeneity of the population are chosen, and different combinations of parameters which represent the relative costs for different sub-populations are tested for simulations. The theoretical analysis in the second part can thus be verified. To summarize the paper, we present some limitations of current research, and point out some possible extensions for future work as well.

\medspace

\section{Vaccination game in the heterogeneous mixing populations}\label{sect3}
\subsection{The mathematical model including population heterogeneity and the predicted attack ratios}
The predicted attack ratio is one the most important factors to influence individuals' choices in vaccination games. The calculations of attack ratios are performed by the formulations of epidemic models and the corresponding final size relations. We now consider the structured population with a distribution of sub-population identified by the variable $\zeta$ which is ranging over the stage space $\Omega$ (see more details in \cite{Bai2016,Brauer2009}). The size of each sub-population is assumed to be 
$N(t,\zeta)$ at the time point $t$. We then make some common assumptions on the contact patterns and suppose one group member in $\zeta$ makes $a(\zeta)$ contacts in unit time. Throughout the paper, we only consider the qualified contacts which are sufficient to transmit infections between individuals. Any contacts with very short periods of time are omitted. We also assume that the fraction of contacts made by one member of group $\zeta$ that is with a member of group $\eta$ is $p(\zeta,\eta)$, and we have
\begin{equation}\label{completeProb}
    \int_{\Omega}p(\zeta,\eta)d\eta = 1.
\end{equation}
The total number of contacts made in unit time by a member of group $\zeta$ with members of group $\eta$ is
\begin{equation*}
a(\zeta)p(\zeta,\eta)N(t,\zeta),
\end{equation*}
and by the balance relation we have
\begin{equation}\label{balanceEqn}
a(\zeta)p(\zeta,\eta)N(t,\zeta) = a(\eta)p(\eta,\zeta)N(t,\eta),
\end{equation}
with $\zeta \neq \eta$. If there is no permanent movement between groups and that there are no disease deaths, it is reasonable to assume that $N(t,\zeta)$ is a constant function $N(\zeta)$ about the time $t$ for all $\zeta$. For sub-population $\zeta$, we let $S(t,\zeta)$ denote the number of susceptible individuals at time $t$ and let $\phi(t,\zeta)$ be the total infectivity at time $t$, defined as the sum of products of the number of infected members with each infection age and the mean infectivity for that infection age. We assume that on average members of the population make a constant number $a$ of contacts in unit time. We next define $B(\tau,\zeta)$ the fraction of infected members remaining infected at infection age $\tau$ and $\pi(\tau,\zeta)$ the mean infectivity at infection age $\tau$, for sub-population $\zeta$. It is straightforward to have
\begin{equation*}
A(\tau) = B(\tau)\pi(\tau),
\end{equation*}
the mean infectivity of members of the population with infection age $\tau$. We assume no disease deaths, so the total population is a constant. Thus, the general age-of-infection model for the structured population is
\begin{equation} \label{general1}
\begin{aligned}
S^{\prime}(t,\zeta) &= -a(\zeta)S(t,\zeta)\int_{\Omega}p(\zeta,\eta)\frac{\phi(t,\eta)}{N(\eta)}d\eta  \\
\phi(t,\zeta) &= \phi_{0}(t,\zeta) + \int_{0}^{\infty}[-S^{\prime}(t-\tau,\zeta)]A(\tau,\zeta)d\tau.
\end{aligned}
\end{equation}
The initial conditions are $S_{0}(\zeta)=(1-p(\zeta))N(\zeta)$ and the vaccine is assumed perfectly effective. 
If the contact pattern is not assumed to be proportionate mixing, it is difficult to obtain the theoretical expression of the basic reproduction number $\mathcal{R}_0$, but we can still formulate the final size relation for model \eqref{general1}, 
\begin{equation}\label{finalsizerelation}
\begin{split}
\log\frac{S(0,\zeta)}{S(\infty,\zeta)} &= \int_{0}^{\infty}a(\zeta)\int_{\Omega}p(\zeta,\eta)\frac{\phi(t,\eta)}{N(\eta)}d\eta dt
\\
&=  \int_{0}^{\infty}a(\zeta)\int_{\Omega}p(\zeta,\eta)\frac{\phi_{0}(t,\eta) + \int_{0}^{\infty}[-S^{\prime}(t-\tau,\eta)]A(\tau,\eta)d\tau}{N(\eta)}d\eta dt
\\
&= a(\zeta)\int_{\Omega}\frac{p(\zeta,\eta)}{N(\eta)}\int_{0}^{\infty}(\phi_{0}(t,\eta) + \int_{0}^{\infty}[-S^{\prime}(t-\tau,\eta)]A(\tau,\eta)d\tau)dtd\eta
\\
&= a(\zeta)\int_{\Omega}\frac{p(\zeta,\eta)}{N(\eta)}\left(\int_{0}^{\infty}\phi_{0}(t,\eta)+\int_{0}^{\infty}\int_{\tau}^{\infty}[-S^{\prime}(t-\tau,\eta)]A(\tau,\eta)dtd\tau\right)d\tau
\\
&=a(\zeta)\int_{\Omega}\frac{p(\zeta,\eta)}{N(\eta)}\left(\int_{0}^{\infty}\phi_{0}(t,\eta)dt + (S(0,\eta)-S(\infty,\eta))\int_{0}^{\infty}A(\eta)dt\right)d\eta.
\end{split}
\end{equation}
It has been calculated in \cite{Brauer2009} that, if the mixing is proportionate, the basic reproduction number is
\begin{equation}\label{R0Pro}
    \mathcal{R}_0 = \int_{\Omega}\int_{0}^\infty p(\eta)a(\eta)A(t,\eta)dtd\eta.
\end{equation}

It is reasonable to consider the more practical way to divide the population into $n (\geq 2)$ finite subgroups (\cite{Brauer2008,Brauer2018,Brauer2019,Bai2020,Bai2021,DelValle2013,Mossong2008}). We thus obtain the simpler notations for the sizes of sub-populations $\alpha_i N$ ($i=1,\cdots,n$) with $0<\alpha_i<1$ and $N$ the total population size. We denote $\mathbb{P}$ the contact pattern matrix, where the element $p_{ij}\geq 0$ represents the fraction of contacts made by a member of subgroup $i$ that is with a member of subgroup $j$. Similar with relation in Equation \eqref{completeProb}, for any $i = 1,\cdots,n$, we have
\begin{equation}\label{constraint1}
	\sum_{j=1}^{n}p_{ij} = 1.
\end{equation}
For each subgroup, we assume that the contact rate is $\beta_i$ for $i=1,\cdots,n$ (\cite{Bai2021,Brauer2008,Jacquez1988,Nold1980}). The discrete version of balance relation equations between different subgroups is
\begin{equation}\label{constraint2}
	\alpha_i\beta_i p_{ij} = \alpha_j\beta_j p_{ji},
\end{equation}
for $i,j=1,\cdots,n$. The contact matrix $\mathbb{P}$ is admissible if and only if the constraints in \eqref{constraint1} and \eqref{constraint2} are satisfied (\cite{Jacquez1988,Busenberg1991}). Various types of contact matrices $\mathbb{P}$ for non-random mixing patterns have been investigated in different epidemic models (\cite{Bai2020,Bai2021,Brauer2008,Jacquez1988,Nold1980,DelValle2013,Feng2015,Cui2018}). One of the important mixing patterns is the proportionate mixing, which suggests that mixing is random but constrained by the activity level (\cite{Bai2020,Bai2021,Brauer2008}). The element in contact matrix $\mathbb{P}$ is obtained,
	\begin{equation*}
		p_{ij} = \frac{\alpha_j \beta_j}{\displaystyle\sum\limits_{j=1}^{n}\alpha_j \beta_j}.
	\end{equation*}
We then simplify the notations to have
\begin{equation*}
p_{1i}=p_{2i}=\cdots=p_{ni}=:p_{i}, \quad \forall i \in \{1,\cdots,n\}.    
\end{equation*}
Similar as in many other previous investigations, we assume the proportionate mixing for the structured population and are therefore able to subsequently simply further analysis. 

\medskip
We propose the discrete version of the general model \eqref{general1},
\begin{equation}\label{general2}
\begin{aligned}
S^{\prime}_i(t) &= -\beta_i S_i(t)\sum_{j=1}^{n}p_{j}\frac{\phi_j(t)}{N_j} \\
\phi^{\prime}_i(t) &= \phi_{0,i}(t) + \int_{0}^{\infty}[-S_i^{\prime}(t-\tau)]A_i(\tau)d\tau.
\end{aligned}    
\end{equation}
It is straightforward to calculate the basic reproduction number to model \eqref{general2},
\begin{equation}\label{R02}
\mathcal{R}_0 = \sum_{i=1}^{n} p_i \beta_i \int_{0}^{\infty} A_i(\tau)d\tau.   
\end{equation}
We then derive the similar final size relation to model \eqref{general2}. For sub-population $i$, we have
\begin{equation}\label{finalsizedisc}
    \log\frac{S_{0,i}}{S_{\infty,i}} = \beta_i \sum_{j=1}^{n}\frac{p_j}{N_j}\left( \int_{0}^{\infty}\phi_{0,j}(t)dt + (S_{0,j}-S_{\infty,j})\int_{0}^\infty A_j(t)dt\right).
\end{equation}
The attack ratios in different sub-populations are defined as the ratio of the number of individuals who have been infected during the course of the epidemic and the total number of individuals in the sub-population. For example, in sub-population $i$, the predicted attack ratio is
\begin{equation*}
    \pi_i = \frac{S_{0,i}-S_{\infty,i}}{S_{0,i}}.
\end{equation*}
The attack ratios $\pi_i$ are functions of vaccine coverage levels in all sub-populations, the effects of vaccine coverage levels on predicted ratios were discussed in \cite{Bai2016,Bai2016a}. Throughout the paper, we assume that the predicted attack ratios $\pi_i$ are monotonically decreasing functions about vaccine coverage levels in sub-populations. 

\medskip

\medskip

\subsection{Vaccination game in the heterogeneous mixing population}
In a heterogeneous mixing population with $2$ sub-groups, the vaccination game is studied by setting the elements of the following payoff matrix in Table \eqref{game1}, based on the fact that, prior to the occurrence
of vaccination, there are exact two strategies for individuals to choose (\cite{Bai2016,Bai2018,Bai2021,Bauch2004,Bai2016a}). Fitness of different strategies can be calculated from this payoff matrix. For simplicity, we assume that the vaccine is perfectly effective, each vaccinated individual is fully protected by the vaccine. For the case of imperfect vaccine, the formulation and its analysis are similar (\cite{Bai2016a,Bai2021}).

The following payoff matrix describes several possible outcomes of the game: first of all, the individual belongs to either subgroup $1$ or subgroup $2$; then the individual is able to chooses strategy "Vaccinated" ($V$) or strategy "Non-vaccinated" ("S"); it is possible that this individual will interact with individuals who choose different strategies. If the individual is in subgroup $1$ and chooses strategy $V$, the payoff is simply the cost of the vaccination $-C_{v1}$. This is true whether this individual is paired with another individual with strategy $V$ or with strategy $S$, because the vaccine is considered fully effective and there is no chance for this individual to get infected. If
this individual is in subgroup $1$ and adopts strategy $S$, there are some different possible outcomes. The first two possibilities are to pair with individuals with strategy $V$. In each of these two cases, since the other individual is fully protected by the vaccine and has no chance to be infected, this individual will not be infected either, which renders the payoff $0$. The other two possible outcomes are to pair with an individual who is in either subgroup $1$ or subgroup $2$ and adopts strategy $S$. If the unvaccinated individual is in subgroup $1$, the probability of being infected is $\pi_1(p,q)$, where $p$ and $q$ are the current vaccine coverage levels in both subgroups. If the cost
of infection is $C_{i1}$, it is straightforward to calculate the payoff of that $S_1$ being paired with another $S_1$ is $-\pi_1(p,q)C_{i1}$. It is convenient to define the relative costs of vaccination for both subgroups,
\begin{equation}\label{relativecosts}
    C_{ri} = \frac{C_{vi}}{C_{ii}}, \quad i=1,2.
\end{equation}
The key assumption throughout our paper is that $C_{r1} \neq C_{r2}$, it yields that the relative costs of vaccination for different subgroups are varied. 
It is reasonable to assume that both $C_{r1}$ and $C_{r2}$ are less than $1$. It is obvious that if the epidemic is more contagious or the vaccine is safe with very few side effects and is with low price, both relative vaccination costs are small. The formal vaccination game is expressed by the payoff matrix in Table \eqref{game1}.

\medspace

\begin{table}[H]
	\centering
	\begin{tabular}{ |p{2cm}||p{2.2cm}|p{2.2cm}|p{2.2cm}|p{2.2cm}|  }	
		\hline
		\multicolumn{5}{|c|}{\textbf{Vaccination game in a population with two subgroups}} \\
		\hline
		& $V_1$ & $S_1$  & $V_2$ & $S_2$\\
		\hline
		$V_1$  & $-C_{v1}$    &  $-C_{v1}$  & $-C_{v1}$  & $-C_{v1}$ \\
		\hline
		$S_1$  & $0$      & $-\pi_{1}(p,q)C_{i1}$  & $0$ & $-\pi_{2}(p,q)C_{i1}$ \\
		\hline
		$V_2$  & $-C_{v2}$ & $-C_{v2}$ & $-C_{v2}$ & $-C_{v2}$ \\
		\hline
		$S_2$  & $0$ & $-\pi_{1}(p,q)C_{i2}$ & $0$ & $-\pi_{2}(p,q)C_{i2}$ \\
		\hline
	\end{tabular}
		\caption{Payoff matrix of the vaccination game in the heterogeneous mixing population with two subgroups. Subscripts $1,2$ represent two subgroups. $p$ and $q$ represent the vaccine coverage levels in two subgroups. It is noted that all individuals can not choose which subgroup they belong to, but can only choose between the adoptions of strategy $V$ or strategy $S$.}
\label{game1}	
\end{table}

\medskip

\section{Predictions of vaccine coverage levels in two subgroups}

Based on the vaccination game which is described in Table \eqref{game1}, we formulate the replicator equations to study the process of decision making regarding whether or not to be vaccinated. Individuals are making decisions based on the nature of the epidemic, the properties of the vaccines and other individuals' decisions. It is noticed that the attack ratios $\pi_{1}(p,q)$ and $\pi_{2}(p,q)$ reach their maximum values when $p=0$ and $q=0$, which is
\begin{equation}\label{max}
\max_{1 \geq p,q \geq 0}\pi_{1,2}(p,q) = \pi_{1,2}(0,0).
\end{equation}
If $\pi_1(0,0) < C_{r1}$, $S$ is the dominant strategy of the game in sub-population $1$. No matter how other individuals play this game, choosing strategy $S$ always has a higher payoff for all individuals in sub-population $1$. This eventually leads to the vaccine coverage level $p= 0$ in this sub-population. Similarly, if $\pi_2(0,0) < C_{r2}$, $S$ is the dominant strategy of the vaccination game in sub-population $2$.

\medskip
At time point $t$, the proportions of individuals who adopt strategy $V$ in sub-population $1$ and sub-population $2$ are $x_{v1}(t)$ and $x_{v2}(t)$, respectively. The proportions of individuals who adopt strategy $S$ in these two sub-populations are $x_{s1}(t)$ and $x_{s2}(t)$, respectively. Thus, the replicator equations are
\begin{equation}\label{2dreplicatorequation}
	\begin{aligned}
		\dot{x_{v1}}(t) &= x_{v1}(t)(f_{v1}(t) - \bar{f_{1}}(t)), \\
		\dot{x_{s1}}(t) &= x_{s1}(t)(f_{s1}(t) - \bar{f_{1}}(t)), \\
		\dot{x_{v2}}(t) &= x_{v2}(t)(f_{v2}(t) - \bar{f_{2}}(t)), \\
		\dot{x_{s2}}(t) &= x_{s2}(t)(f_{s2}(t) - \bar{f_{2}}(t)).
	\end{aligned}
\end{equation}
Where $\bar{f_{i}}(t)$ denotes the average fitness/payoffs for individuals in sub-population $i$ at time $t$, we have
\begin{equation}\label{payoff}
\begin{aligned}
\bar{f_1} &= (1-p)f_{s1} + p f_{v1}, \\
\bar{f_2} &= (1-q)f_{s2} + q f_{v2}.
\end{aligned}    
\end{equation}
The payoffs for vaccinated individuals and non-vaccinated individuals in both sub-populations can be calculated accordingly,
\begin{equation}\label{fitness1}
	\begin{aligned}
		f_{v1} &= -C_{v1}, \\
f_{s1} &= -\pi_{1}(p,q)C_{i1}p_1(1-p) - \pi_{2}(p,q)C_{i1}p_{2}(1-q), \\
		f_{v2} &= -C_{v2}, \\
f_{s2} &= -\pi_{1}(p,q)C_{i2}p_1(1-p) - \pi_{2}(p,q)C_{i2}p_{2}(1-q).
\end{aligned}	
\end{equation}
Because the proportionate mixing is assumed, we are able to simplify the contact probabilities as $p_1 := p_{11}=p_{21}$ and $p_{2}:= p_{12}=p_{22}$. $p_1$ and $p_2$ are defined to represent the probabilities that the contact is made with an individual in subgroups $1$ and $2$, respectively. We then substitute relations into Equation \eqref{payoff} and Equation \eqref{fitness1} to obtain
\begin{equation}\label{2dreplicatorequationModified}
	\begin{aligned}
		\dot{x_{v1}} &= x_{v1}(1-p)(f_{v1} - f_{s1}), \\
		\dot{x_{s1}} &= x_{s1}p(f_{s1} - f_{v1}), \\
		\dot{x_{v2}} &= x_{v2}(1-q)(f_{v2} - f_{s2}), \\
		\dot{x_{s2}} &= x_{s2}q(f_{s2} - f_{v2}).
	\end{aligned}
\end{equation}
It is noticed that the fractions of both sub-populations are fixed, which are
\begin{equation*}
    x_{v1}+x_{s1}=1, \quad x_{v2}+x_{s2}=1.
\end{equation*}
We subsequently simplify the replicator equations in \eqref{2dreplicatorequationModified} as the following system,
\begin{equation}\label{2dreplicatorequationFinal}
	\begin{aligned}
		p^{\prime} &= p(1-p)(f_{v1} - f_{s1}), \\
		q^{\prime} &= q(1-q)(f_{v2} - f_{s2}),
  \end{aligned}
\end{equation}
where $p$ and $q$ the vaccine coverage levels in two sub-populations and are functions about the time. $f_{v1}$, $f_{s1}$, $f_{v2}$ and $f_{s2}$ are calculated in Equation \eqref{fitness1}. It is computed that there are $8$ candidates of evolutionary stable strategies for system \eqref{2dreplicatorequationFinal},
\begin{equation}\label{eqicand}
\begin{pmatrix}
p \\
q
\end{pmatrix} = \Bigg\{\begin{pmatrix}
0 \\
0
\end{pmatrix},\begin{pmatrix}
1 \\
1
\end{pmatrix} ,\begin{pmatrix}
0 \\
1
\end{pmatrix} ,\begin{pmatrix}
1 \\
0
\end{pmatrix} ,\begin{pmatrix}
0 \\
q^\#(0)
\end{pmatrix} ,\begin{pmatrix}
1 \\
q^\#(1)
\end{pmatrix} ,\begin{pmatrix}
p^\ast(0) \\
0
\end{pmatrix} ,\begin{pmatrix}
p^\ast(1) \\
1
\end{pmatrix} \Bigg\}.   
\end{equation}
We apply the notation $q^\#(0)$ to represent the value of $q$ when $p=0$ and satisfies the following condition
\begin{equation*}
    \pi_1(0,q^\#(0))p_1 + \pi_2(0,q^\#(0))p_2(1-q^\#(0)) = C_{r2}.
\end{equation*}
The notation $q^\#(1)$ represents the value of $q$ when $p=1$ and satisfies the condition
\begin{equation*}
    \pi_1(1,q^\#(1))p_1 + \pi_2(1,q^\#(1))p_2(1-q^\#(1)) = C_{r2}.
\end{equation*}
The notations $p^\ast(0)$ and $p^\ast(1)$ can be interpreted in the same way for $p$, when $q$ has the fixed values of $0$ and $1$, respectively. The following lemma states the uniqueness of $q^\#(0)$, $q^\#(1)$, $p^\ast(0)$ and $p^\ast(1)$.

\medskip

\begin{lemma}
For the fixed value of $0 \leq q \leq 1$, there exists at most one value of  $p$ such that
\begin{equation*}
\pi_{1}(p,q)p_1(1-p) + \pi_{2}(p,q)p_{2}(1-q) = C_{r1}.
\end{equation*}
\end{lemma}

\medskip
\begin{proof}
Suppose $q$ is fixed, the left side of the above equation is a function about $p$. We differentiate the function about $p$ to obtain the first order derivative
\begin{equation*}
    \dfrac{\partial \pi_1(p,q)}{\partial p}p_1(1-p) - \pi_1(p,q)p_1 + \dfrac{\partial \pi_2(p,q)}{\partial p}p_2(1-q). 
\end{equation*}
It is observed that both $\dfrac{\partial \pi_1(p,q)}{\partial p}$ and $\dfrac{\partial \pi_2(p,q)}{\partial p}$ are negative, so the first order derivative is strictly negative. It yields that the left side function is a monotonic decreasing function and thus there exists at most one value of $p$ to satisfy the condition for any fixed value of $q$.
\end{proof}
It is similar to prove that for the fixed value of $0 \leq p \leq 1$, there exists at most one value of $q$ such that
\begin{equation*}
\pi_{1}(p,q)p_1(1-p) + \pi_{2}(p,q)p_{2}(1-q) = C_{r2}.
\end{equation*}

It is worth mentioning that, for some cases, some or all of these $4$ equilibria may not exist.

\medskip

In order to examine the local stability of these equilibrium points, we linearize the system \eqref{2dreplicatorequationFinal} near each equilibrium and compute the corresponding Jacobian matrix $J$. $4$ elements of the Jacobian matrix $J$ are computed as,
\begin{equation}\label{Jacobian}
\begin{aligned}
J_{11} = & C_{i1}\bigg((1-2p)(\pi_1 p_1 (1-p) + \pi_2 p_2 (1-q) - C_{r1}) 
          +p(1-p)(\frac{\partial \pi_1}{\partial p}p_1 (1-p)+\frac{\partial \pi_2}{\partial p}p_2 (1-q)-\pi_1 p_1)\bigg),   \\
J_{12} = & C_{i1}p(1-p)\bigg(\frac{\partial \pi_1}{\partial q}p_1 (1-p)+\frac{\partial \pi_2}{\partial q}p_2 (1-q)-\pi_2 p_2\bigg), \\ 
J_{21} = & C_{i2}q(1-q)\bigg(\frac{\partial \pi_1}{\partial p}p_1 (1-p)+\frac{\partial \pi_2}{\partial p}p_2 (1-q)-\pi_1 p_1\bigg), \\
J_{22} = & C_{i2}\bigg((1-2q)(\pi_1 p_1 (1-p) + \pi_2 p_2 (1-q) - C_{r2}) 
          +q(1-q)(\frac{\partial \pi_1}{\partial q}p_1 (1-p)+\frac{\partial \pi_2}{\partial q}p_2 (1-q)-\pi_2 p_2)\bigg).
\end{aligned}
\end{equation}
The trace of the Jacobian matrix can be expressed as
\begin{equation}
\begin{aligned}
 & J_{11} + J_{22} \\
              = &  C_{i1}\bigg((1-2p)(\pi_1 p_1 (1-p) + \pi_2 p_2 (1-q) - C_{r1})  +p(1-p)(\frac{\partial \pi_1}{\partial p}p_1 (1-p)+\frac{\partial \pi_2}{\partial p}p_2 (1-q)-\pi_1 p_1)\bigg)   \\
         +&  C_{i2}\bigg((1-2q)(\pi_1 p_1 (1-p) + \pi_2 p_2 (1-q) - C_{r2})  +q(1-q)(\frac{\partial \pi_1}{\partial q}p_1 (1-p)+\frac{\partial \pi_2}{\partial q}p_2 (1-q)-\pi_2 p_2)\bigg).
         \end{aligned}    
\end{equation}
The expression of the determinant of the Jacobian matrix is
\begin{equation}
\begin{aligned}
& J_{11}J_{22} - J_{12}J_{21} \\
                    = & C_{i1}C_{i2}\Bigg[(1-2p)(1-2q)\bigg(\pi_1 p_1 (1-p) + \pi_2 p_2 (1-q) -C_{r1}\bigg)\bigg(\pi_1 p_1 (1-p) + \pi_2 p_2 (1-q) -C_{r2}\bigg) \\
                 & + (1-2p)q(1-q)\bigg(\pi_1 p_1 (1-p) + \pi_2 p_2 (1-q) -C_{r1}\bigg)\bigg(\frac{\partial \pi_1}{\partial q} p_1 (1-p) + \frac{\partial \pi_2}{\partial q} p_2 (1-q) -\pi_2 p_2\bigg) \\
                  & + (1-2q)p(1-p)\bigg(\pi_1 p_1 (1-p) + \pi_2 p_2 (1-q) -C_{r2}\bigg)\bigg(\frac{\partial \pi_1}{\partial p} p_1 (1-p) + \frac{\partial \pi_2}{\partial p} p_2 (1-q) -\pi_1 p_1\bigg) \Bigg].
\end{aligned}    
\end{equation}

\medskip
The trace and the determinant of the Jacobian matrix imply the positiveness of the real parts of all eigenvalues of the matrix, and further yields the local stability conditions near the candidate equilibrium points. We use the following theorem to summarize the partial results.

\medskip
\begin{theorem}
The equilibria $
\begin{pmatrix}
0 \\
0
\end{pmatrix},
\begin{pmatrix}
1 \\
1
\end{pmatrix}$ are locally unstable for the system of replicator equations \ref{2dreplicatorequationFinal}.
\end{theorem}

\medskip
It can be calculated that near the equilibria $
\begin{pmatrix}
0 \\
0
\end{pmatrix},
\begin{pmatrix}
1 \\
1
\end{pmatrix}$, both the trace and the determinant of the Jacobian matrix are positive, therefore these two equilibira are locally unstable. However, for the rest of the candidates of equilibria, the computations of the traces and the determinants are based on the choices of values of $p$ and $q$. The stability conditions of other equilibira are to be determined. The indication of the theorem is to exclude the possibilities that both sub-populations choose exact the same strategy when facing the potential outbreak of the epidemic. It is unlikely for the structured population to be vaccinated (or non-vaccinated) entirely if all individuals in the population are making decisions for maximizing personal payoffs. It is also impossible to have the overall vaccine coverage level be $0$.

\medskip

\begin{remark}
One special case of the considered vaccination game in the two-subgroup population is that the relative costs for two sub-populations are equal. It has been investigated in \cite{Bai2021} that, besides the candidates of equilibria we considered, there exists more equilibrium $(p^\ast, q^\ast)$ satisfying the condition
\begin{equation*}
    \pi_1(p^\ast,q^\ast)p_1(1-p^\ast) + \pi_2(p^\ast,q^\ast)p_2(1-q^\ast) = C_{r}, 
\end{equation*}
where $C_r := C_{r1} = C_{r2}$. 
\end{remark}

\medskip

\begin{remark}
There are at most $8$ Nash equilibria for the vaccination game in the heterogeneous mixing population with two sub-groups, if the relative vaccination costs are varied. However, if the relative vaccination costs are identical for both subgroups, there exist infinitely many Nash equilibira. 
\end{remark}

\medskip

\section{Numerical simulations}
\begin{figure}[H]
	\centering
	\includegraphics[width = 1\textwidth]{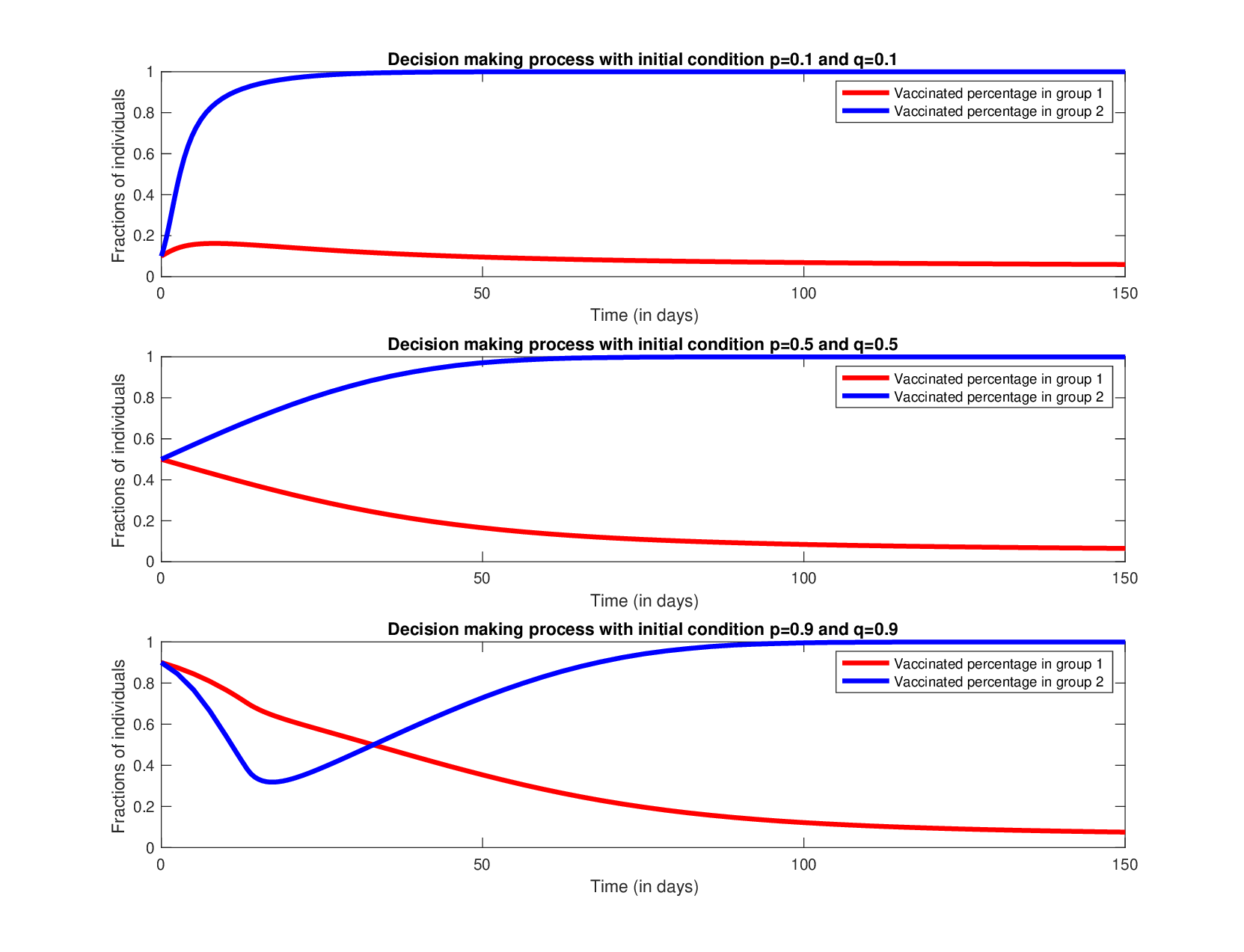}
	\caption{Relative costs are $1/5$ and $1/10$ for two sub-populations, respectively.}
	\label{fig:exp1}
\end{figure}

In this section, we perform a series of numerical simulations to verify the theoretical results and show the decision making process in the population regarding whether to be vaccinated, prior to the outbreak of an epidemic. In order to simplify the numerical simulations, we make the simple assumption that the infectious period is exponentially distributed. The general epidemic model \eqref{general2} with integro-differential equations can be reduced to the simple ordinary differential equation (ODE) system. The reduced ODE system is
\begin{equation}\label{modelODE}
	\begin{cases}
		S_{i}^{\prime} &= -S_{i}\beta_{i}\sum\limits_{j=1}^{2}p_{j}\frac{I_j}{N_j} \\
		I_{i}^{\prime} &= S_{i}\beta_{i}\sum\limits_{j=1}^{2}p_{j}\frac{I_j}{N_j} -  \gamma I_{i} \\
		R_{i}^{\prime} &= \gamma I_{i}.
	\end{cases}
\end{equation}
Where $\gamma$ reflects the averaged recovery rate for both sub-populations. The final size relations can be derived from Equation \eqref{finalsizedisc}. In the next step, we set the key demographic parameters for the population sizes with the following values,
\begin{equation*}
N=4.5\times10^6, \quad N_1 = 5\times10^5, \quad N_2 = 4\times10^6.
\end{equation*}
The epidemiological parameters for recovery rate and the activity levels are 
\begin{equation*}
\gamma = 0.2, \quad \beta_1 = 1, \quad \beta_2 = 0.25.    
\end{equation*}
The parameters which are related with the contact pattern in the population can thus be calculated,
\begin{equation*}
p_{1} = \frac{1}{3}, \quad p_2 = \frac{2}{3}.   
\end{equation*}
It yields that, with the probability of $\displaystyle\frac{1}{3}$, a randomly chosen individual in the population will encounter another individual from sub-population $1$. We then calculate the basic reproduction number for model \eqref{modelODE} is $\mathcal{R}_0=2.5$. It is worth mentioning that although the size of sub-population $1$ is much smaller than the size of sub-population $2$, because of the significant higher level of activity, we consider sub-population $1$ is the major contributor to the transmission of the epidemic among the whole population. For detailed discussions regarding the major/minor infection contributors, we refer to \cite{Bai2021}.
It was investigated that the relative vaccination costs for both sub-populations are the key factors in the vaccination decision making process, we therefore choose different combinations of $C_{r1}$ and $C_{r2}$ to simulate the final outcome of the decision making process. For each set of numerical experiments, we test for three initial conditions $10\%$, $50\%$ $90\%$, which represent that the initial attitudes in the two sub-populations towards vaccination are very negative, almost neutral and very positive. We first consider the following combination of parameters which represents the costs of infection, the costs of vaccination and the relative costs for the two sub-populations, respectively,
\begin{equation}
\begin{aligned}
&C_{i1} = 0.5, \quad C_{v1} = 0.1, \quad C_{r1} = 1/5. \\
&C_{i2} = 2, \quad C_{v2} = 0.2, \quad C_{r2} = 1/10.
\end{aligned}    
\end{equation}
The relative cost of vaccination for sub-population $1$ is greater than that for sub-population $2$, which yields that the strategy of to be vaccinated is the preferred one for sub-population $2$. It is therefore reasonable to expect that the vaccine coverage level in sup-population $2$ is higher than the coverage level in sub-population $1$. The simulation of the decision making process is presented in Figure \ref{fig:exp1}. The final vaccination coverage levels in two sub-populations are
\begin{equation*}
p^\ast=5.2\%, \quad q^\ast=100\%.
\end{equation*}
We notice that the final predicted vaccine coverage in sub-population $1$ is significantly low. It is also interesting to point out that, as shown in the third sub-figure of Figure \ref{fig:exp1}, if the initial willingness of being vaccinated is strong in each sup-population, there will be a turning point as time evolves. During the early stage of the decision-making process, the fractions of individuals who are willing to be vaccinated are both high. Subsequently individuals in both sub-populations keep trying to maximize their own payoffs and speculate that the high vaccine coverage from the own group and another group would provide indirect protection against the infection. The reasonable strategies for both groups are to switch to non-vaccinated and these will result in the decreases of the vaccine coverage levels in both sub-populations. As the fractions of individuals who are willing to take the vaccine continue to decline, individuals in sub-population $2$ consider the fact that the vaccination relative cost is lower for them and evaluate the risk of infection is actually much higher. Then the strategy of being vaccinated is again becoming favored and this will lead to the increase of vaccine coverage level in this sub-population. The increase of possible vaccine coverage level in sub-population $2$ directly causes the continued decline of possible vaccination level in sub-population $1$. When the decision making process finishes, the vaccine coverage level in sub-population $2$ reaches $100\%$ and the vaccine coverage level in sub-population $1$ reaches to a very low level. 
\medskip

\medskip

\begin{figure}[H]
	\centering
	\includegraphics[width = 1\textwidth]{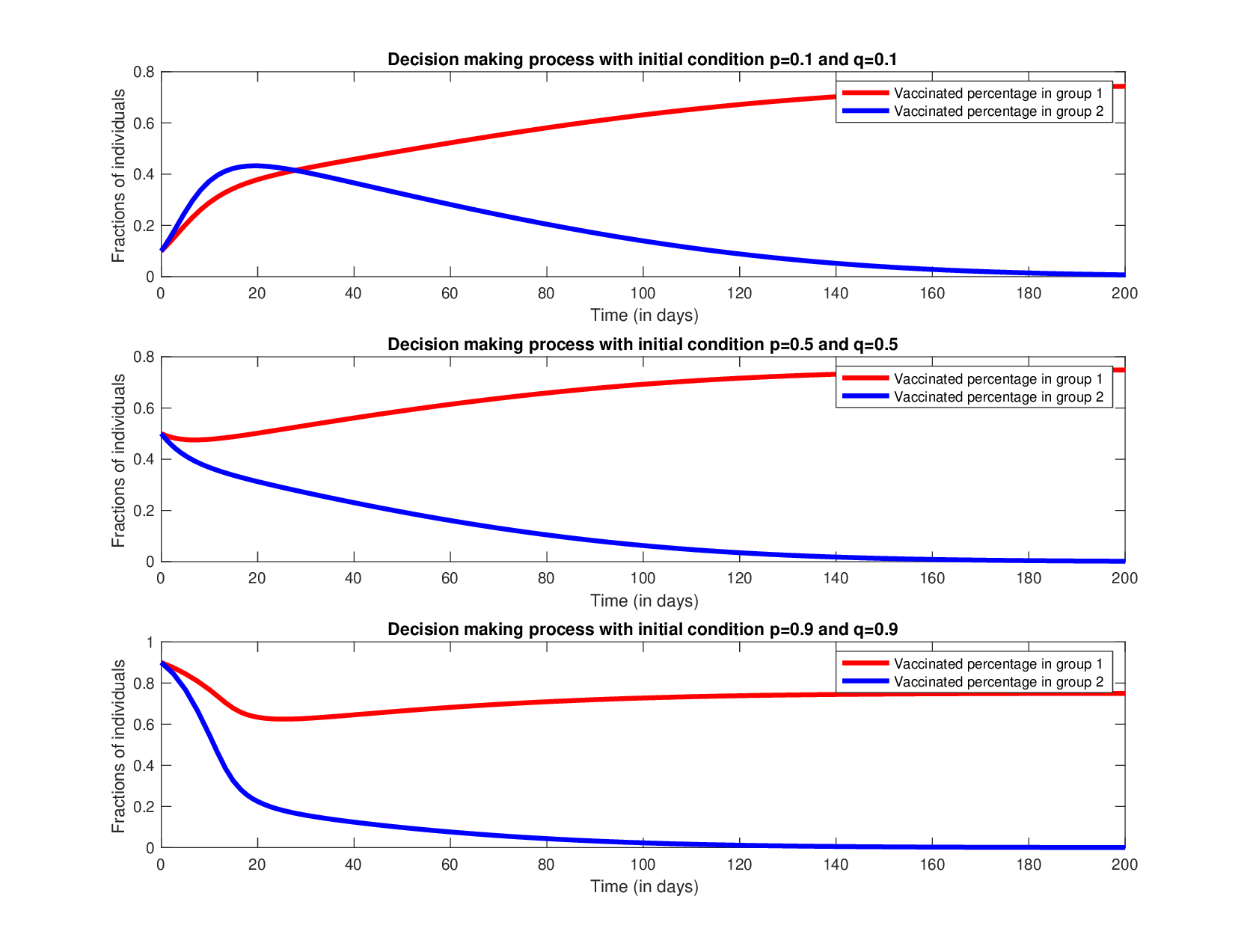}
	\caption{Relative costs are $1/5$ and $1/4$ for two sub-populations, respectively.}
	\label{fig:exp2}
\end{figure}

We next only change the value of $C_{v2}$ to $0.5$ in parameter values, which yields the relative cost for sub-population $2$ becomes $1/4$. The simulation result of the decision making process is presented in Figure \ref{fig:exp2}. It is shown that due to the relatively higher cost of vaccination, individuals in sub-population $2$ eventually choose the strategy of non-vaccinated. We also observe the turning point in the first sub-figure of Figure \ref{fig:exp2}, which initially both sub-populations have the low degree of willingness regarding vaccination. The final vaccination coverage levels for two-populations are computed,
\begin{equation*}
   p^\ast = 75\%, \quad q^\ast = 0. 
\end{equation*}

\medskip

\medskip

\begin{figure}[H]
	\centering
	\includegraphics[width = 1\textwidth]{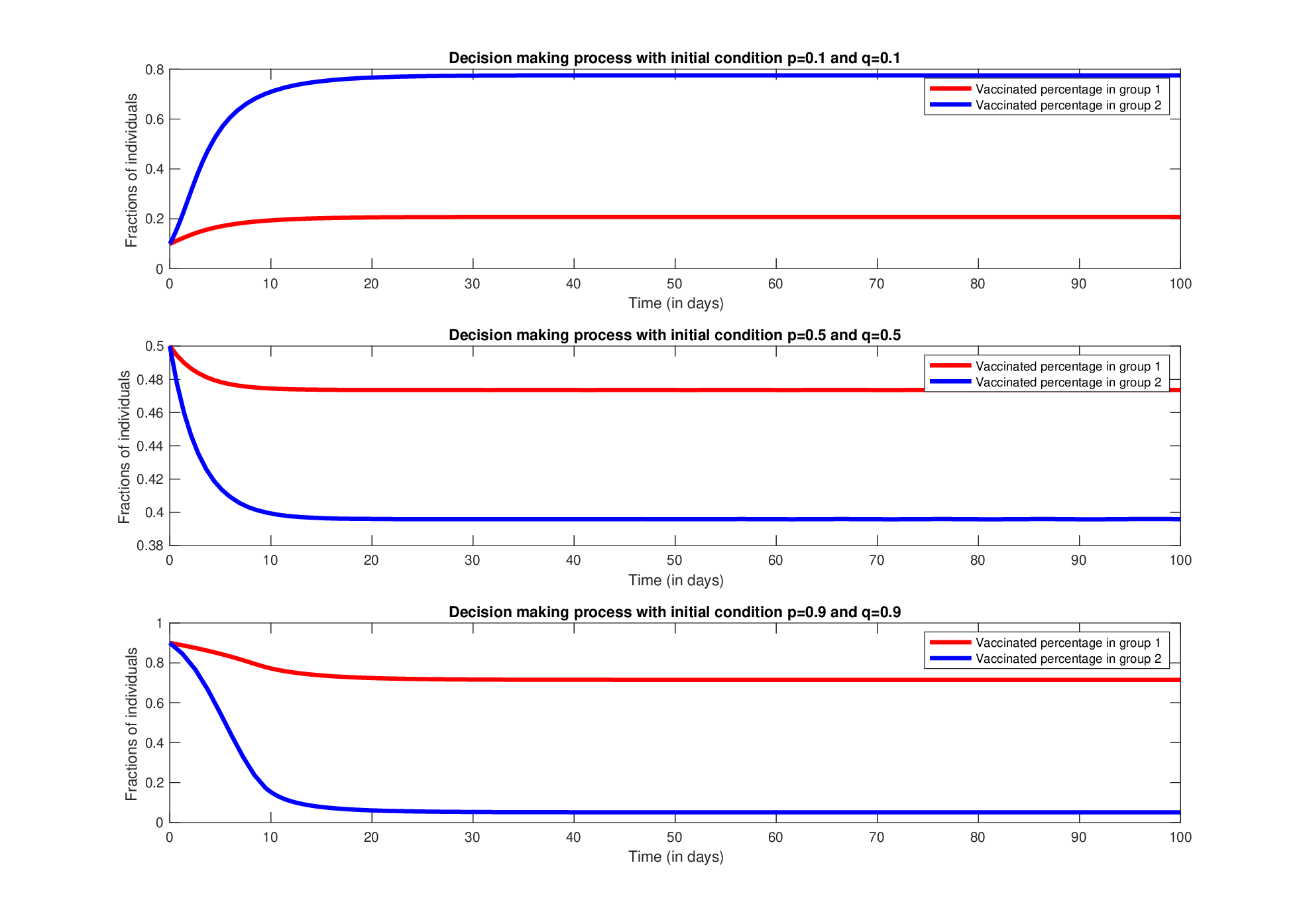}
	\caption{Relative costs are both $1/5$ for two sub-populations.}
	\label{fig:exp3}
\end{figure}

The third experiment is to change the cost of vaccination for sub-population $2$ to $0.4$, which results in the case that the relative vaccination costs for both sub-populations are equal. The simulation of the decision making process is presented in Figure \ref{fig:exp3}. It is shown that, for this vaccination game, there exist multiple stable states. If different initial conditions are considered, the dynamical system will evolve and eventually converge to different stable states. The first sub-figure in Figure \ref{fig:exp3} represents the condition that the initial willingness towards vaccination are both low in two sub-groups, which are $10\%$. As time evolves, both percentages increase and reach the stable state
\begin{equation*}
p^\ast = 20.8\%, \quad q^\ast = 77.5\%.    
\end{equation*}
If the initial willingness are both $50\%$ in two sub-populations, the final vaccination levels for two sub-populations are
\begin{equation*}
p^\ast = 47.3\%, \quad q^\ast = 39.6\%.    
\end{equation*}
If the initial willingness are both $90\%$ in two sub-populations, the final vaccination levels for two sub-populations are
\begin{equation*}
p^\ast = 71.4\%, \quad q^\ast = 5.1\%.    
\end{equation*}

We thus are able to draw the conclusion that, for the case of equal relative costs, it is more difficult to accurately predict the final vaccine coverage level for the decision making process. The final outcome is highly sensitive to the initial conditions for all the sub-populations.

\medskip

\medspace

\section{Conclusion and future work}
We have formulated mathematical models in the structured population to investigate the impact of heterogeneity on the predicted vaccine coverage levels. The complete modeling framework essentially consists of two parts: the age-of-infection model and the game theoretical formulation. The age-of-infection was formulated mainly to describe the potential outcome of the epidemic in the whole population for the given vaccine coverage levels, the ultimate goal is to derive the attack ratios or the final size relations from the model. The attack ratios in sub-populations were further applied in the vaccination games as a key ingredient. The vaccination games were subsequently formulated to simulate the decision making process in the structured population. We then applied the technique of replicator equations to analyze the time-dependent dynamical system and were able to compute the candidates of all evolutionary stable strategies (ESS). It is noticed that the uniqueness of Nash equilibrium for such vaccination games does not hold due to the inclusion of population heterogeneity. Only partial results on the stability of the stable strategies were obtained, because of the complexities of the computations of the Jacobian matrices near the equilibria. If the relative vaccination costs are considered in the structured population, there are finitely many candidates of evolutionary stable strategies. If the relative vaccination costs are equal for all sub-populations, we have infinitely many candidates of evolutionary stable strategies. We then performed several groups of numerical experiments to verify the analytical results. The inclusion of population heterogeneity does not level up the overall vaccine coverage level to reach the herd immunity, it even complicates the prediction of vaccine coverage for all sub-populations. 

There are many other types of population heterogeneity to be investigated in this topic. For example, we have made the simple assumption that the distributions of infectious periods are identical for all sub-populations in order to reduce the system of integro-differential equations to the system of ordinary differential equations. This can be improved by introducing the heterogeneity from different distributions of infectious periods in sub-populations. We speculate that this extension will not affect the game theoretical analysis, but will change the computations of attack ratios in different sub-populations. More advanced numerical methods need to be applied for simulations, such as the collocation method or the finite difference/element method (\cite{Breda2020,Messina2023}).

\medspace

\ \\ 
\noindent {\bf Declaration of Competing Interest:}
None.

\medspace

\noindent {\bf Acknowledgements:}
The work was supported in part by the BNU-HKBU UIC startup research grant UICR$0700037$-$22$ and by the Guangdong Provincial Key Laboratory of Interdisciplinary Research and Application for Data Science, BNU-HKBU United International College, project code $2022$B$1212010006$  and UIC research grant R$0400001$-$22$.

\clearpage
\newpage

\clearpage
\newpage

\bibliography{bibfile}
\bibliographystyle{plain}
\end{document}